\documentclass[%
reprint,
superscriptaddress,
amsmath,amssymb,
prb,
]{revtex4-2}

\usepackage{graphicx}
\usepackage{dcolumn}
\usepackage{bm}
\usepackage[colorlinks,urlcolor=blue,linkcolor=blue,citecolor=blue,anchorcolor=blue]{hyperref}
\usepackage{color}

\begin{document}
	
	
	\title{Floquet topological insulators with hybrid edges}
	
	\author{Boquan Ren}
	\affiliation{Key Laboratory for Physical Electronics and Devices of the Ministry of Education \& Shaanxi Key Lab of Information Photonic Technique, School of Electronic Science and Engineering, Faculty of Electronic and Information Engineering, Xi'an Jiaotong University, Xi'an 710049, China}
	
	\author{Yaroslav V. Kartashov}
	\affiliation{Institute of Spectroscopy, Russian Academy of Sciences, Troitsk, Moscow, 108840, Russia}
	
	\author{Hongguang Wang}
	\affiliation{Key Laboratory for Physical Electronics and Devices of the Ministry of Education \& Shaanxi Key Lab of Information Photonic Technique, School of Electronic Science and Engineering, Faculty of Electronic and Information Engineering, Xi'an Jiaotong University, Xi'an 710049, China}
	
	\author{Yongdong Li}
	\affiliation{Key Laboratory for Physical Electronics and Devices of the Ministry of Education \& Shaanxi Key Lab of Information Photonic Technique, School of Electronic Science and Engineering, Faculty of Electronic and Information Engineering, Xi'an Jiaotong University, Xi'an 710049, China}
	
	\author{Yiqi Zhang}
	\email{zhangyiqi@xjtu.edu.cn}
	\affiliation{Key Laboratory for Physical Electronics and Devices of the Ministry of Education \& Shaanxi Key Lab of Information Photonic Technique, School of Electronic Science and Engineering, Faculty of Electronic and Information Engineering, Xi'an Jiaotong University, Xi'an 710049, China}
	
	\date{\today}
	
	\begin{abstract}
		\noindent
		Topological edge states form at the edges of periodic materials with specific degeneracies in their modal spectra, such as Dirac points, under the action of effects breaking certain symmetries of the system. In particular, in Floquet topological insulators unidirectional edge states appear upon breakup of the effective time-reversal symmetry due to dynamical modulations of the underlying lattice potential. However, such states are usually reported for certain simple lattice terminations, for example, at zigzag or bearded edges in honeycomb lattices. Here we show that unconventional topological edge states may exist in Floquet insulators based on arrays of helical waveguides with hybrid edges involving alternating zigzag and armchair segments, even if the latter are long. Such edge states appear in the largest part of the first Brillouin zone and show topological protection upon passage through the defects. Topological states at hybrid edges persist in the presence of focusing nonlinearity of the material. Our results can be extended to other lattice types and physical systems, they lift some of the constraints connected with lattice terminations that may not support edge states in the absence of effects breaking time-reversal symmetry of the system and expand the variety of geometrical shapes in which topological insulators can be constructed.
	\end{abstract}
	
	\maketitle
	

	\section{Introduction}
	
	Topological insulators are considered as a new state of matter, and their most representative feature is the existence of topologically protected states at their edges that are responsible for transport of excitations despite insulating bulk~\cite{hasan.rmp.82.3045.2010, qi.rmp.83.1057.2011}. The phenomenology of topological insulators originally developed in solid-state physics, has been extended to a wide variety of photonic~\cite{lu.np.8.821.2014, ozawa.rmp.91.015006.2019, kim.lsa.9.130.2020, smirnova.apr.7.021306.2020, ota.nano.9.547.2020, leykam.nano.9.4473.2020, segev.nano.10.425.2021, parto.nano.10.403.2021, yan.aom.2001739.2021, tang.lpr.1.2100300.2022, rechtsman.nature.496.196.2013, haldane.prl.100.013904.2008, wang.nature.461.772.2009, lindner.np.7.490.2011, hafezi.np.7.907.2011, stuetzer.nature.560.461.2018, yang.nature.565.622.2019, mukherjee.science.368.856.2020, maczewsky.science.370.701.2020, yang.light.9.128.2020} and other physical systems~\cite{susstrunk.science.349.47.2015, huber.np.12.621.2016, yang.prl.114.114301.2015, he.np.12.1124.2016, peng.nc.7.13368.2016, lu.np.13.369.2017, zhang.cp.1.97.2018, ma.nrp.1.281.2019, jotzu.nature.515.237.2014, goldman.pnas.110.6736.2013, nalitov.prl.114.116401.2015, jean.np.11.651.2017, klembt.nature.562.552.2018, albert.prl.114.173902.2015, hadad.ne.1.178.2018, imhof.np.14.925.2018, olekhno.nc.11.1436.2020, helbig.np.16.747.2020, li.nsr.8.nwaa1192.2021}. Photonic topological insulators are most frequently constructed on periodic artificial materials, such as honeycomb~\cite{rechtsman.nature.496.196.2013, noh.prl.120.063902.2018, wu.nc.8.1304.2017,ablowitz.pra.96.043868.2017}, kagome~\cite{ablowitz.pra.99.033821.2019, zhong.rip.12.996.2019, ivanov.pra.103.053507.2021}, Lieb~\cite{bandres.cleo.2014, li.prb.97.081103.2018, ablowitz.pra.99.033821.2019, ivanov.ol.45.1459.2020} or other lattices possessing specific degeneracies in their spectra, such as Dirac points~\cite{leykam.aipx.1.101.2016}. Edge states then form with eigenvalues within topological bandgap that may open between these points under the action of different effects.
	
	\begin{figure*}[htpb]
		\centering
		\includegraphics[width=\textwidth]{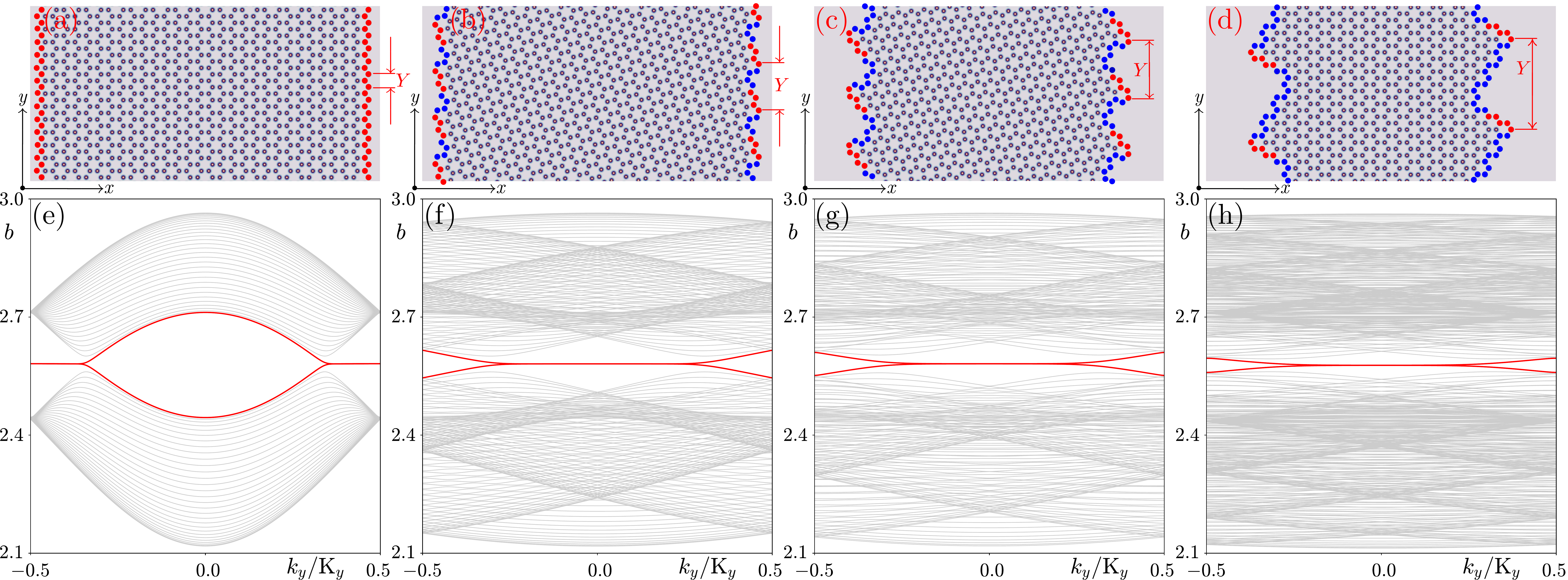}
		\caption{(a-d) Examples of honeycomb arrays with hybrid edges that are periodic along the $y$ axis with different periods $Y$ and are truncated along the $x$-axis. Waveguides belonging to zigzag and armchair edge segments are indicated by the red and blue dots, respectively. (a) Lattice with usual zigzag boundary, (b-d) lattices containing 1, 2, or 4 armchair segments on each period (called below case-1, case-2, and case-4 ribbons, respectively). (e-h) Corresponding band structures at $r_0=0$ in the first Brillouin zone $-0.5{\rm K}_y \le k_y \le 0.5{\rm K}_y$, where levels giving rise to edge states are highlighted by the red color, while bulk states are shown gray. Lattices are shown in the window: $-40\le x\le 40$ and $-20\le y\le 20$.}
		\label{fig1}
	\end{figure*}
	
	Typically edge states are reported for specific lattice terminations (edge geometries) that also determine the location of their eigenvalues in the first Brillouin zone (BZ) in projected spectra of such truncated structures. For example, honeycomb lattices that were used for demonstration of many topological phenomena~\cite{rechtsman.prl.111.103901.2013, plotnik.nm.13.57.2014, song.nc.6.6272.2015, zhang.nc.11.1902.2020} may have simple zigzag, bearded, and armchair edges. The majority of works on topological edge states in such lattices addresses first two types of lattice terminations~\cite{plotnik.nm.13.57.2014, kohmoto.prb.76.205402.2007}, zigzag or bearded, because in the absence of effects breaking time-reversal symmetry, these terminations lead to the appearance of nontopological edge states in wide intervals of Bloch momenta. The situation is somewhat different for armchair terminations because in the nontopological regime they do not lead to appreciable existence intervals of the edge states. Nevertheless, the very possibility of the formation of the topological edge states at the armchair edges in Floquet systems was mentioned in~\cite{rechtsman.nature.496.196.2013}, and the propagation of linear and nonlinear edge states on the armchair edges of Floquet structure was studied in~\cite{ablowitz.ol.40.4635.2015}. Still, the properties of linear and nonlinear edge states in topological insulators with mixed edges, involving zigzag and armchair terminations of different variable widths were not studied yet. Similar situation is also encountered in well-known kagome and Lieb lattices~\cite{zhong.rip.12.996.2019, li.prb.97.081103.2018} and for various photonic topological platforms, including helical waveguide arrays~\cite{rechtsman.nature.496.196.2013, lumer.prl.111.243905.2013, titum.prl.114.056801.2015, leykam.prl.117.013902.2016, leykam.prl.117.143901.2016, bandres.prx.6.011016.2016, maczewsky.nc.8.13756.2017, mukherjee.nc.8.13918.2017, ablowitz.pra.99.033821.2019, ivanov.apl.4.126101.2019, mukherjee.science.368.856.2020, yang.light.9.128.2020, ivanov.acs.7.735.2020} that receive considerable attention these days, which well mimic the Floquet mechanism in time domain~\cite{kitagawa.prb.82.235114.2010, rudner.prx.3.031005.2013, rudner.nrp.2.229.2020}, see also works~\cite{zhang.acs.4.2250.2017,stuetzer.nature.560.461.2018, lustig.nature.567.356.2019, he.nc.10.4194.2019, unal.prr.1.022003.2019, schuster.prl.123.266803.2019,unal.prl.125.053601.2020, esin.sa.6.eaay4922.2020, afzal.prl.124.253601.2020, zhu.prb.103.L041402.2021, pyrialakos.nm.21.634.2022, yin.elight.2.8.2022, nagulu.ne.5.300.2022} for other promising results obtained on Floquet insulator platforms. The investigation of topological insulators with mixed terminations including segments that do not support nontopological edge states is therefore important because it may allow to substantially extend the possibilities for construction of topological circuits and devices and lift the restriction on their potential geometrical shapes (by removing the constrains connected with ``nontopological terminations" that would lead to loss of confinement for excitations passing through them).
	
	In this article, we propose a new type of the photonic Floquet topological insulator based on the array of helical waveguides with \textit{hybrid} edges involving combined zigzag and armchair segments. Topological unidirectional edge states form and show clear topological protection in such structures even if the length of the armchair segments, which do not support edge states in the absence of waveguide rotation (at least in the tight-binding limit), notably exceeds that of the zigzag segments. Surprisingly, we found that edge states supported by such hybrid terminations in ribbon geometry appear in substantially larger fraction of the BZ in projected spectrum in comparison with usual Floquet insulators with pure zigzag edges. Our Floquet system with broken time-reversal symmetry is qualitatively different from previously studied graphene nanoribbons with hybrid edges supporting the formation of the ``end'' states described by the Su-Schrieffer-Heeger model~\cite{groning.nature.560.209.2018, li.nc.12.5538.2021}.
	
	We also found that topological edge states at \textit{hybrid} edges persist in the nonlinear regime and that such edges can support propagation of localized topological edge solitons. Indeed, nonlinearity in topological systems can lead to a number of intriguing phenomena ranging from modulational instability~\cite{kartashov.optica.3.1228.2016} and bistability of the edge states~\cite{kartashov.prl.119.253904.2017, zhang.lpr.13.1900198.2019, zhang.pra.99.053836.2019}, topological transitions~\cite{maczewsky.science.370.701.2020, xia.light.9.147.2020, hadad.ne.1.178.2018, zangeneh.prl.123.053902.2019}, to formation of topological solitons in the bulk~\cite{lumer.prl.111.243905.2013, mukherjee.science.368.856.2020} or at the edge~\cite{ablowitz.pra.90.023813.2014,leykam.prl.117.143901.2016, ablowitz.pra.96.043868.2017, ivanov.acs.7.735.2020, ivanov.ol.45.1459.2020, zhong.ap.3.056001.2021, ren.nano.10.3559.2021, kirsch.np.17.995.2021, tang.chaos.161.112364.2022} of the insulator. Here we analyze stability of the nonlinear edge states at hybrid edges and illustrate the possibility of edge soliton formation.
	
	\section{Linear edge states}
	\subsection{Band structures and edge states}
	
	Light propagation in helical waveguide arrays can be described by the nonlinear Schr\"odinger-like equation with focusing cubic nonlinearity
	\begin{equation}\label{eq1}
		i \frac{\partial \psi}{\partial z}=-\frac{1}{2} \left( \frac{\partial^2}{\partial x^2} + \frac{\partial^2}{\partial x^2} \right) \psi
		-\mathcal{R}(x,y,z) \psi-|\psi|^{2} \psi,
	\end{equation}
	where $\psi$ is the dimensionless complex amplitude of the field, $x$ and $y$ are normalized transverse coordinates, $z$ is the normalized propagation distance, and the function 
	$\mathcal{R}(x,y,z)=\mathcal{R}(x,y,z+Z)$ 
	describes $z$-periodic refractive index profile with longitudinal period $Z$. The array consists of identical helical waveguides of width $\sigma$ placed in the nodes $(x_m,y_n)$ of the honeycomb grid 
	\begin{equation}
		\mathcal{R}(x,y,z) = p \sum_{mn} e^{ -\frac{[x-x_m-r_0 \sin(\omega z)]^2}{\sigma^2} - \frac{[y-y_n-r_0\cos(\omega z)+r_0]^2}{\sigma^2 }}, 
	\end{equation}
	where $p$ is the array depth and $\omega=2\pi/Z$. We adopt the parameters $d=1.7$ ($17\, \mu \rm m$ spacing between waveguides), $r_0 \sim 0.0-1.0$ (helix radius up to $10\, \mu \rm m$), $p=11$ (refractive index modulation depth $\delta n \sim 1.2\times 10^{-3}$), $\sigma=0.4$ (4 $\mu \rm m$-wide waveguides), and $Z=8$ (helix period $\sim 9.1\, \rm mm$) typical for fs-laser written arrays at $\lambda=800\,\rm nm$~\cite{kirsch.np.17.995.2021, kartashov.prl.128.093901.2022, tan.ap.3.024002.2021, li.ap.4.024002.2022}. We consider ribbons with hybrid edges truncated along the $x$-axis and periodic in $y$: $\mathcal{R}(x,y,z)=\mathcal{R}(x,y+Y,z)$. 
	
	First of all, it is necessary to look at the band structure and edge states with different terminations 
	by neglecting the nonlinear term in {\bf Equation}~(\ref{fig1}).
	{\bf Figure}~\ref{fig1} shows honeycomb lattice with usual zigzag edges [Figure~\ref{fig1}(a)] and structures with hybrid zigzag-armchair terminations for progressively increasing lengths of the armchair segments (increasing $Y$ periods) [Figure~\ref{fig1}(b,c,d)]. We refer to lattices in Figure~\ref{fig1}(b,c,d) as to case-1, case-2, and case-4 ribbons, in accordance with the number of armchair segments on one $y$-period. Notice that the length of the armchair segments can substantially exceed that of the zigzag segments [Figure~\ref{fig1}(d)].
	
	\begin{figure}[htpb]
		\centering
		\includegraphics[width=0.9\columnwidth]{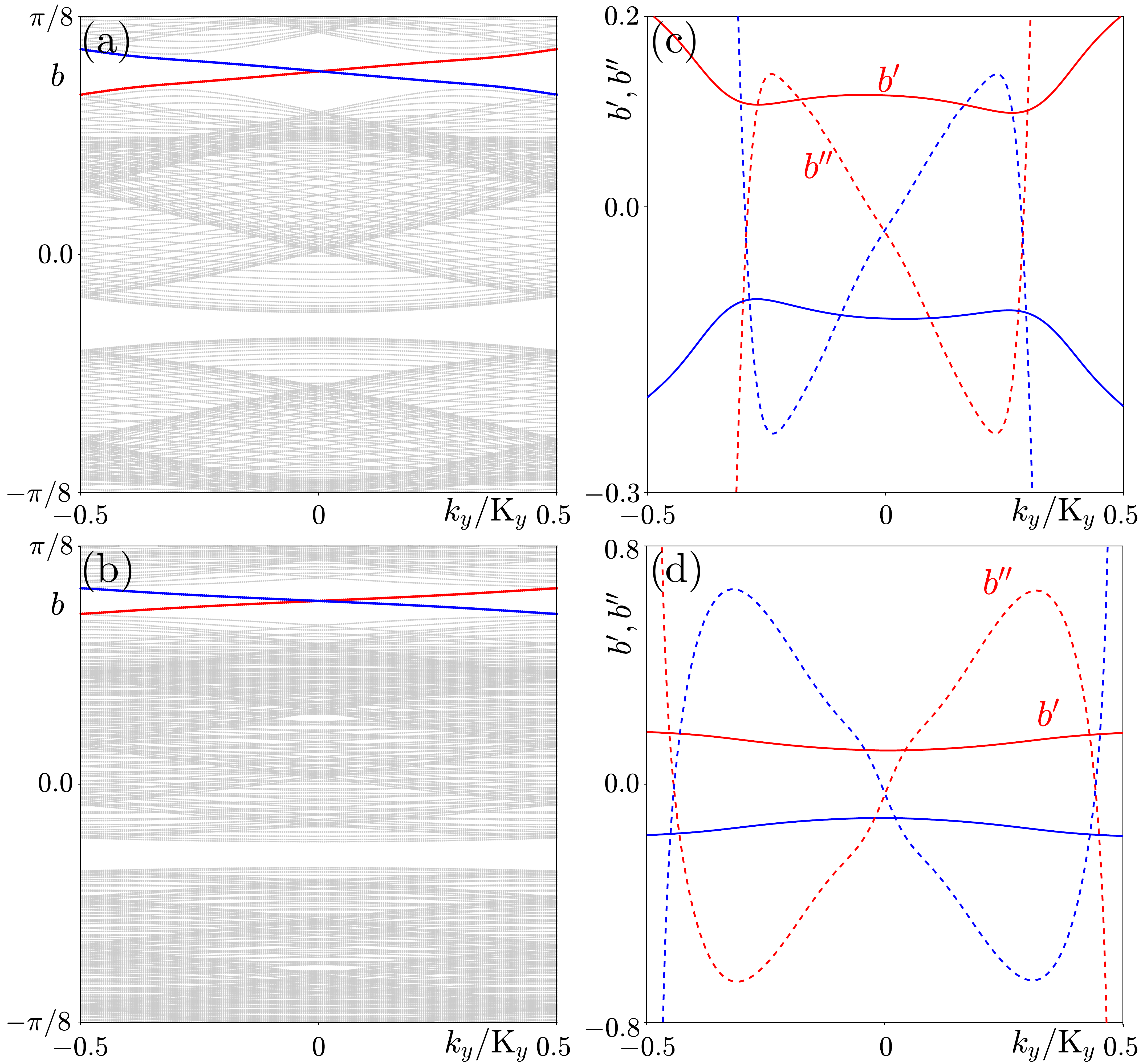}
		\caption{Quasi-propagation constant spectrum and derivatives of the quasi-propagation constants for linear edge states versus $k_y/\textrm{K}_y$ for case-1 (a),(c) and case-4 (b),(d) ribbons at $r_0=0.5$ and $Z=8$. Levels giving rise to topological edge states are shown in red and blue, while bulk states are shown gray. Red (blue) edge states are localized on the right (left) edge of the ribbon. Derivatives $b'=db/dk_y$ are shown with solid lines, while derivatives $b''=d^2 b/dk_y^2$ are shown with dashed lines [line colors correspond to colors used in (a),(b)].}
		\label{fig2}
	\end{figure}
	
	For ribbons with helical waveguides the linear eigenmodes are the Floquet-Bloch waves 
	\begin{equation}
		\psi(x,y,z)=\phi_{k_y}(x,y,z) e^{ibz}=u_{k_y}(x,y,z) e^{ik_yy+ibz}, 
	\end{equation}
	where $b=b(k_y)$ is the quasi-propagation constant (similar to quasi-energy in driven quantum systems~\cite{kitagawa.prb.82.235114.2010,rudner.nrp.2.229.2020}) within the first longitudinal BZ $b\in [-\omega/2, \omega/2)$ with $\omega=2\pi/Z$, $k_y$ is the Bloch momentum within the first transverse BZ $k_y\in[-{\rm K}_y/2,{\rm K}_y/2)$, where $\textrm{K}_y=2\pi/Y$. The function $u_{k_y}(x,y,z)=u_{k_y}(x,y+Y,z)=u_{k_y}(x,y,z+Z)$ is periodic both in $y$ and $z$. When  $r_0=0$ and waveguides are straight, $b$ becomes usual propagation constant. Such linear ``static" band structures $b(k_y)$ for ribbons from Figure~\ref{fig1}(a)-(d) are presented in Figure~\ref{fig1}(e)-(h). The levels that give rise to edge states are shown red (we mark them in the entire BZ for simplicity, even though edge states exist in limited $k_y$ intervals), while bulk states are shown gray. In clear contrast to lattice with zigzag boundaries with localized states emerging near the border of the BZ [Figure~\ref{fig1}(e)], in structures with hybrid boundaries two degenerate edge states appear in a broad region in the center of the BZ [Figure~\ref{fig1}(f)-(h)]. Waveguide rotation opens topological gap, lifts the degeneracy of the edge states, and makes them unidirectional, transforming the system into Floquet topological insulator. This is seen from band structures presented for $r_0 \ne 0$ in Figure~\ref{fig2}(a) and \ref{fig2}(b), where red/blue curves correspond to topological edge states localized at the right/left edges. 
	Corresponding derivative $b'=db/dk_y$ [solid lines in Figure~\ref{fig2}(c) and \ref{fig2}(d)] determines group velocity of the topological edge states via $v=-b'$, while dispersive properties are determined by $b''=d^2b/dk_y^2$ [dashed lines in Figure~\ref{fig2}(c) and \ref{fig2}(d)]. One can see that states from blue branch move in the positive $y$ direction, while those on the red branch move in the negative $y$ direction. It should be mentioned that topological properties of honeycomb lattices of helical waveguides (characterizing topology of its bulk bands) are well-documented~\cite{rechtsman.nature.496.196.2013} and can be described by proper generalization of the Chern number for Floquet systems \cite{kitagawa.prb.82.235114.2010, rudner.nrp.2.229.2020}. The existence of the topological edge states on hybrid edges is thus consistent with bulk-edge correspondence principle.
	
	\begin{figure}[htpb]
		\centering
		\includegraphics[width=\columnwidth]{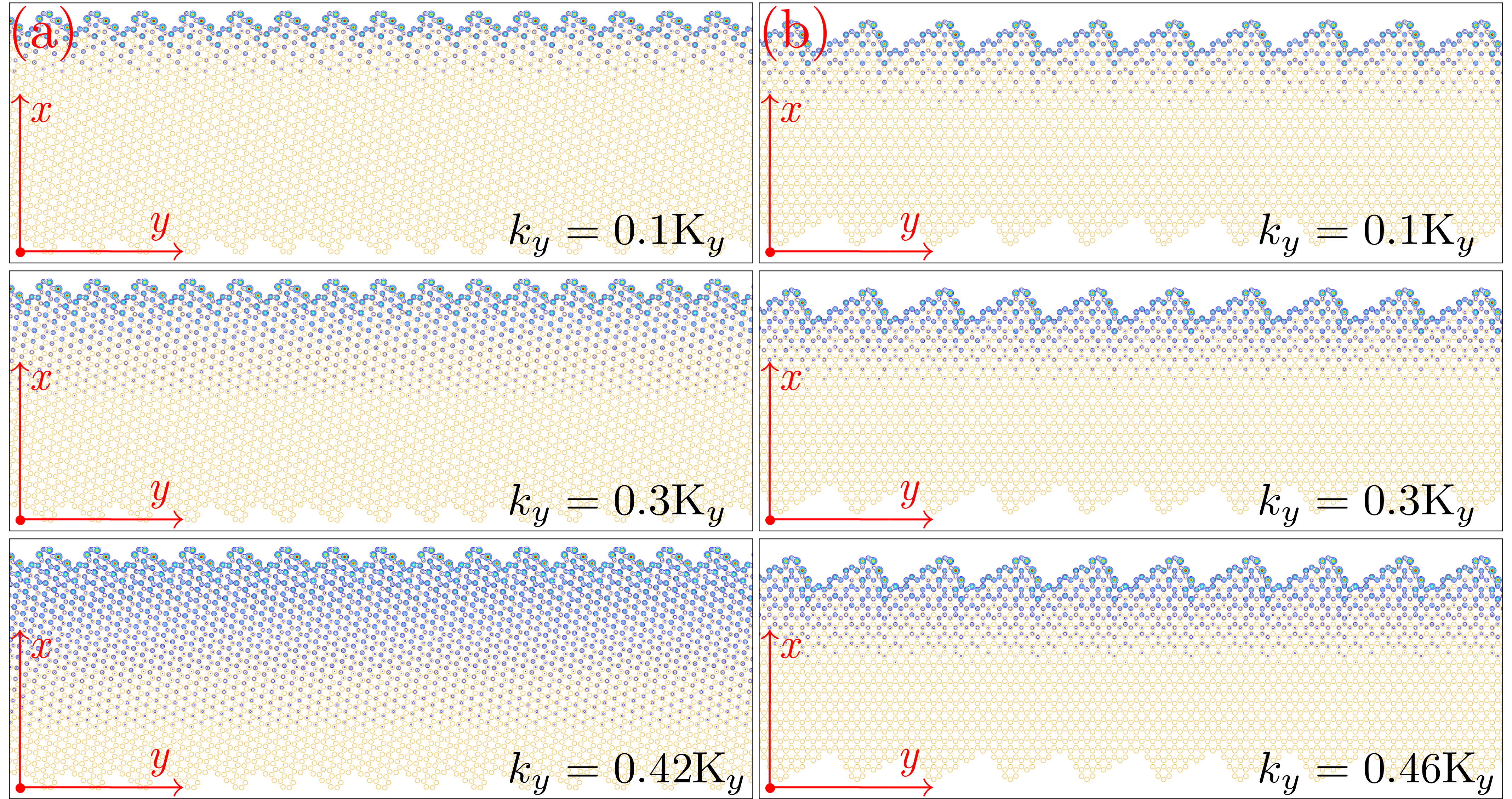}
		\caption{Field modulus distributions at $z=0$ in linear topological edge states with different momenta $k_y$ for case-2 (a) and case-4 (b) ribbons with $r_0=0.5$, $Z=8$. Orange circles represent the lattice sites. All panels are shown in the window: $-35\le x\le 35$ and $-100\le y\le 100$.}
		\label{fig3}
	\end{figure}
	
	Representative $|\psi|$ distributions at $z=0$ in Floquet-Bloch states localized at hybrid edges are presented in Figure~\ref{fig3}. Here we choose case-2 [Figure~\ref{fig3}(a)] and case-4 [Figure~\ref{fig3}(b)] ribbons as exemplary structures. The remarkable property of the system with hybrid edges is that localized edge states are encountered in much larger fraction of the BZ in comparison with system with pure zigzag or bearded edges. The best edge state localization is achieved near the BZ center at $k_y=0$, while towards the border of the BZ localization becomes weaker and edge states gradually transform into bulk ones (for case-1, case-2, and case-4 ribbons delocalization for selected parameters occurs for $k_y > 0.36{\rm K}_y$, $k_y>0.43{\rm K}_y$, and $k_y>0.46{\rm K}_y$, respectively). Thus, the existence range of topological edge states surprisingly increases with increasing length of the armchair segments. This also leads to enhancement of edge state localization at fixed $k_y$, as seen from comparison of states in case-2 and case-4 ribbons.
	
	\subsection{Propagation dynamics}
	
\begin{figure}[htpb]
		\centering
		\includegraphics[width=\columnwidth]{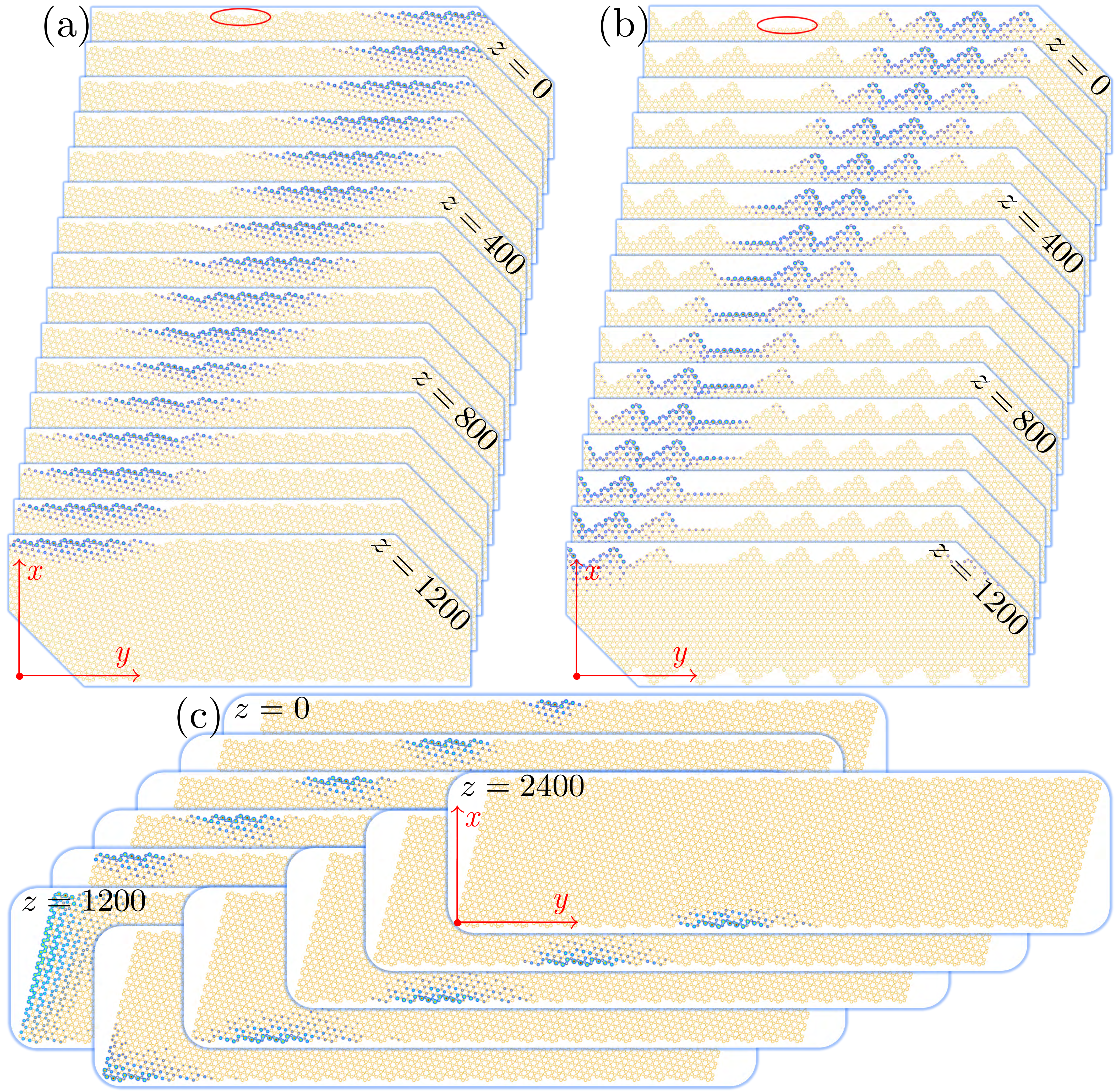}
		\caption{(a) Field modulus distributions at different distances illustrating propagation of the linear edge state with $k_y=0.1{\rm K}_y$ in case-1 ribbon with a defect on the edge indicated by the red ellipse. (b) Setup is as (a), but for case-4 ribbon with larger defect. (c) Circulation of topological edge soliton obtained in the presence of nonlinearity at $b_{\rm nl}=0.002$, $k_y=0.1{\rm K}_y$, $b''=-0.1214$, $\chi=0.1358$ in case-1 ribbon (top and bottom edges are hybrid edges, while left and right edges are purely armchair). Other parameters are as in Figure~\ref{fig2}.}
		\label{fig4}
	\end{figure}

	One prominent property of the edge states from Figure~\ref{fig3} is that they can bypass defects or disorder without reflection due to their topological protection. We choose topological edge states with $k_y=0.1{\rm K}_y$ in case-1 and case-4 ribbons for illustration that this property holds in structures with hybrid edges. To illustrate unidirectional propagation we superimpose a wide Gaussian envelope on such states. Some waveguides were removed in case-1 and case-4 ribbons to create defects, as shown by the red ellipses in Figure~\ref{fig4}(a) and \ref{fig4}(b) (in the latter case we removed not just single guide, but the whole ``tooth" on the edge). Field modulus distributions at different distances in Figure~\ref{fig4}(a,b) clearly illustrate topological protection and absence of backward/bulk radiation.
	
	\begin{figure*}[htpb]
		\centering
		\includegraphics[width=\textwidth]{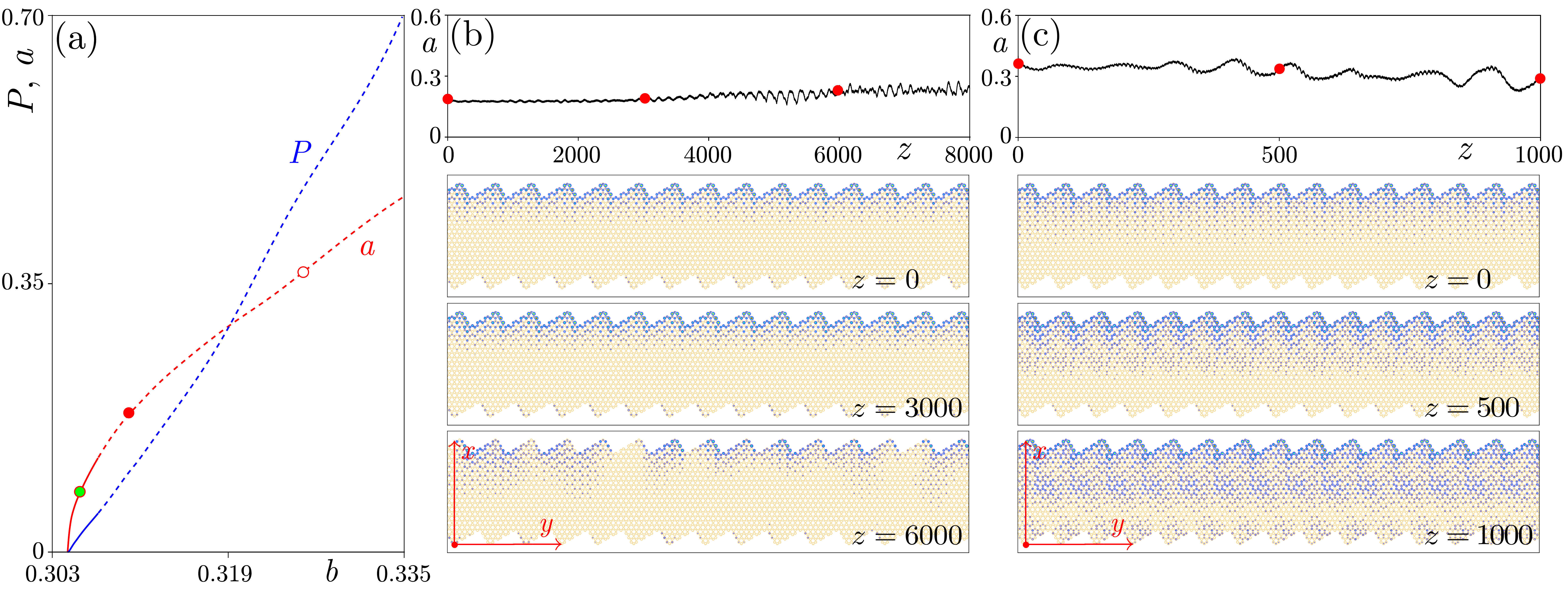}
		\caption{(a) Nonlinear edge state family in case-4 ribbon at $k_y=0.1\textrm{K}_y$. Peak amplitude $a$ (red curve) and power $P$ per period of the edge state as functions of nonlinear generalization of quasi-propagation constant $b$. The green, red, and white dots correspond to $b=0.3067$ ($P=0.02$), $b=0.311$ ($P=0.1$), and $b=0.327$ ($P=0.475$), respectively. Solid and dashed lines represent metastable and unstable states. (b) Peak amplitude $a$ of the perturbed edge state corresponding to the red dot in (a) versus distance $z$ and associated $|\psi|$ distributions at different propagation distances. (c) Setup is as (b), but corresponding to the white dot in (a). All states are shown within the window $-35\le x\le 35$ and $-150\le y\le 150$.}
		\label{fig5}
	\end{figure*}
	
	\section{Nonlinear edge states and edge solitons}
	
	In addition to linear topological edge states, our system with hybrid edges also supports nonlinear edge states and topological edge solitons. In Figure~\ref{fig5} we show representative families of the nonlinear Floquet-Bloch modes bifurcating from linear edge states in case-4 ribbon. Their profiles are modified by the nonlinearity that also leads to a shift of the quasi-propagation constant. These families were obtained using iterative method of~\cite{lumer.prl.111.243905.2013}. The procedure of calculation of the nonlinear Floquet-Bloch modes can be divided into several steps. (1) We propagate the linear edge state $\psi_e^{\rm in}$ with a given power $P$ (that will determine eventually quasi-propagation constant of the nonlinear Floquet-Bloch state) according to Eq.~(\ref{eq1}) to obtain the dynamical lattice modified by the nonlinearity, i.e. $\mathcal{R}_e=\mathcal{R}+|\psi_e^{\rm in}|^2$. (2) After this we propagate all linear eigenstates $\psi^{\rm in}_{n\in N}$ of $\mathcal{R}$ that include $\psi_e^{\rm in}$ in the modified dynamical lattice $\mathcal{R}_e$ for a whole period $Z$ and obtain corresponding output distributions $\psi^{\rm out}_{n\in N}$, where $N$ is the number of eigenstates. (3) One calculates the projection $U_{mn} = \langle \psi^{\rm in}_m, \psi^{\rm out}_n \rangle$, whose eigenvalues are Floquet exponents $e^{iZb_n}$. (4) For each $b_n$, one finds the index $\ell_n$ of the maximum element of the corresponding eigenvector $V_n$. Eigenstates $\psi_{n\in N}^{\rm re}$ of $\mathcal{R}_e$ can then be constructed as $\psi_n^{\rm re}=\sum_{q=1}^N \psi_q^{\rm in} V_q(\ell_n)$. (5) One picks out the modified edge state $\psi_e^{\rm re}$ from $\psi_{n\in N}^{\rm re}$ and normalizes it to the given power $P$. (6) The steps (1)-(5) are repeated until the difference between $\psi_e^{\rm re}$ and $\psi_e^{\rm in}$ reduces below required small level.
	
	Peak amplitude $a$ and power per period 
	\[
	P=\int_{-\infty}^{\infty} dx \int_0^Y |\psi|^2 dy
	\]
	of the nonlinear edge state increase with increasing nonlinear shift of quasi-propagation constant $b$ from its linear value $b=0.3057$ at $k_y=0.1{\rm K}_y$ [Figure~\ref{fig5}(a)]. One can see from this figure that the procedure described above allows us to trace the nonlinear family directly from its bifurcation point from linear edge state, so that one can be sure that this is continuous family of the simplest thresholdless nonlinear solutions. Of course, this does not exclude the possibility of the existence of more complex nonlinear Floquet-Bloch states, but they were not found here. Stability of such nonlinear Floquet-Bloch states was investigated by superimposing random input noise on them with amplitude up to $0.05a$ and propagating them over long distances $z \sim 10^4$. Thus, nonlinear edge state indicated by the green dot ($b=0.3067$ and $P=0.02$) in Figure~\ref{fig5}(a) is metastable and does not decay even at $z=10^4$, while those indicated by the red dot ($b=0.311$ and $P=0.1$) and the white dot ($b=0.327$ and $P=0.475$) are unstable. The dashed lines in Fig.~\ref{fig5}(a) represent unstable branch of the nonlinear edge states, which roughly starts from the case with $P=0.05$ ($b=0.3084$).
	Typical scenario of instability development is illustrated in Figure~\ref{fig5}(b), where one can see that peak amplitude of the edge state starts changing notably only at distances $z>3000$, while wave fragmentation along the edge on developed stage of modulational instability becomes clear at $z \sim 6000$. Nonlinear edge states with higher amplitudes may show somewhat different instability scenario resulting in considerable radiation into the bulk without obvious fragmentation along the edge [see Figure~\ref{fig5}(c) illustrating propagation of perturbed edge state corresponding to the white dot from Figure~\ref{fig5}(a)].
	
	The possibility of development of modulational instability indicates that Floquet insulators with hybrid edges can support localized edge solitons travelling along the edge~\cite{kartashov.optica.3.1228.2016, zhong.ap.3.056001.2021}. We illustrate them here for case-1 ribbon. Such solitons can be constructed as envelope solitons 
	\begin{equation}
		\psi(x,y,z)=A(\eta,z)\phi_{k_y}(x,y,z) e^{ibz}
	\end{equation}
	on the Floquet edge state $\phi_{k_y}(x,y,z)$ with proper sign of the dispersion coefficient $b''$. The equation 
	\begin{equation}
		i \frac{\partial A}{\partial z} = \frac{b''}{2} \frac{\partial^2A}{\partial\eta^2} - \chi \lvert A \rvert^2 A,
	\end{equation}
	where $\eta=y+b'z$, for slowly varying envelope $A(\eta,z)$ of such states can be derived from Equation~(\ref{eq1}) using the method developed in~\cite{ivanov.acs.7.735.2020}. In this equation 
	\begin{equation}
		\chi=\frac{1}{Z}\int_0^Z dz \int_{-\infty}^{+\infty} dx \int_0^Y \lvert \phi_{k_y}(x,y,z) \rvert^4 dy
	\end{equation}
	is the effective period-averaged nonlinearity coefficient. 
	Soliton envelopes existing at $b''<0$ are of the form 
	\begin{equation}
		A(\eta,z) = \sqrt{\frac{2 b_\textrm{nl}}{\chi}} \textrm{sech} \left[ \sqrt{\frac{-2 b_\textrm{nl}}{b''}} (y+b' z) \right] e^{ib_{\rm nl} z},
	\end{equation}
	where $b_{\rm nl}$ is the nonlinearity-induced shift of propagation constant that should be sufficiently small to ensure that slowly varying envelope covers many periods of the ribbon. The example of propagation of such edge soliton with $b_{\rm nl}=0.002$ is presented in Figure~\ref{fig4}(c). This soliton bifurcates from edge state with $k_y=0.1{\rm K}_y$.
	We consider soliton circulation in closed geometry, when case-1 ribbon is made finite also in the $y$ direction [Figure~\ref{fig4}(c)]. One can see that soliton initially slightly broadens and then starts moving as a robust object along the hybrid edge. It broadens at purely armchair edge (around $z=1200$), but after it passes onto opposite hybrid edge of the ribbon, nonlinearity helps it to recombine again into steadily propagating localized state (see the distribution at $z=2400$). Radiation into the bulk is practically absent upon such circulation. Numerical simulations reveal that Floquet edge solitons in structures with hybrid edges survive even after many roundtrips.
	The animations corresponding to Figure\,\ref{fig4}(a)-\ref{fig4}(c) are provided,
	in which the peak amplitude of the soliton $a=\max\{|\psi|\}$ is recorded and displayed simultaneously.
	The animation of the propagation of this soliton in finite structure is up to distances $z\sim20000$ (i.e., 2500 longitudinal periods).
	The animations can be found in \href{https://gr.xjtu.edu.cn/documents/137794/302175/animation_fig4a.mp4/9450af2f-f4d9-a177-97f5-8cbc5bdfac20?t=1670400547642}{Animation 1},
	\href{https://gr.xjtu.edu.cn/documents/137794/302175/animation_fig4b.mp4/b5c7efd5-a058-c7e3-8add-b94ed9024385?t=1670400574504}{Animation 2},
	and \href{https://gr.xjtu.edu.cn/documents/137794/302175/animation_fig4c.mp4/cd919441-7fdd-46eb-9d94-b78b12f71ad8?t=1670400615946}{Animation 3}.
	
	\section{Conclusions}
	
	Summarizing, we have shown that Floquet topological insulators may support topologically protected edge states for previously unexplored hybrid types of lattice terminations, that combine mixed armchair and zigzag segments, even when the former segments constitute larger fraction of the edge. Moreover, such hybrid edges support long-living travelling edge solitons that show robust circulation along the structure periphery. These results can be extended to other types of lattices and various physical systems, where Floquet insulators can be realized, including atomic and optoelectronic systems. They will also allow to expand the variety of geometrical shapes in which topological insulators can be constructed and analyzed.

\noindent\textbf{Acknowledgements}\\
This work was supported by the National Natural Science Foundation of China (Grant Nos.: 12074308, U1537210),
Russian Science Foundation (Grant No.: 21-12-00096), and the
Fundamental Research Funds for the Central Universities
(Grant No.: xzy022022058).
\\

\noindent\textbf{Data availability}\\
The data that supports the results within this paper are available from the corresponding
authors upon reasonable request.
\\

\noindent\textbf{Code availability}\\
The codes used to obtain the results described in this paper are available from the corresponding authors upon reasonable request.
\\

\noindent\textbf{Competing interests}
The authors declare no competing interests.
\\

\noindent\textbf{Correspondence} and requests for materials should be addressed to Y.Z.
\\

%

\end{document}